# How ‘Foundational’ Are Current Molecular Foundation Models?

Francesca Grisoni[1*]

[1]Eindhoven University of Technology, Department of Biomedical Engineering, Institute of Complex Molecular Systems (ICMS), Eindhoven, The Netherlands.
*E-mail: f.grisoni@tue.nl

**Abstract**

Large-scale models have permeated the molecular sciences, yet what makes a model ‘foundational’ in this domain remains poorly defined. This paper proposes three testable criteria for assessing the foundational nature of molecular models: (i) *generality* across molecular entities, properties, and tasks; (ii) *transferability* to new applications with no or minimal task-specific retraining; and (iii) *generalization* beyond the training distribution. Applying these criteria to the state of the art reveals promising progress, particularly visible in biomolecular structure prediction and machine-learned interatomic potentials, although none of the approaches examined fully satisfies all three. Success is concentrated in domains where target properties are consistently defined and training data are abundant, with low label noise relative to physically meaningful variation. More broadly, progress in the molecular sciences appears to depend less on model scale alone than on the quality and structure of available data, as well as the incorporation of prior knowledge into models, prediction tasks, or downstream applications. This work shifts the notion of a molecular foundation model from a descriptive label to a testable hypothesis, offering a framework for assessing current models and guiding future developments.

## Introduction

In the past few years, deep learning – based on neural networks with multiple processing layers[1] – has undergone a paradigm shift thanks to models that are trained on broad data at scale and are adaptable to a wide range of downstream tasks.[2] This shift builds on the ability of deep neural networks to learn 'data-driven' representations, together with empirical evidence that performance can improve as data, model size, and computational resources are scaled up.[3] Notable examples are models like BERT[4] and GPT-3,[5] pretrained once on large text corpora and subsequently adapted to tasks like translation, summarization, and question answering. The term 'foundation model'[2] was recently coined to describe models with such scale and adaptation characteristics.

Deep learning has also driven remarkable advances across the molecular sciences.[6–8] AlphaFold2[9] achieved unprecedented accuracy in protein structure prediction, a breakthrough recognized by the 2024 Nobel Prize in Chemistry.[10] AlphaFold3[11] extended these capabilities to complexes spanning proteins, nucleic acids, and small-molecule ligands. Boltz-2[12] combined structural accuracy[13] with binding-affinity prediction in a single framework. ESM-3,[14] trained on protein sequence, structure, and function, generated a functional fluorescent protein substantially different from known ones. In materials discovery, GNoME[15] proposed hundreds of thousands of candidate stable inorganic compounds, although independent analyses have questioned their novelty and the reliability of the associated synthesis.[16,17] The current advances illustrate the expanding capabilities of molecular machine learning and the growing ambition to develop increasingly general and scalable models across the molecular sciences.

However, the use of the term 'foundation model' has been increasingly adopted across scientific disciplines, often without sufficient justification.[18] This paper therefore provides a conceptual framework for assessing the foundational capabilities of molecular machine learning models, aiming to identify current strengths, limitations, and opportunities for future development. The analysis focuses on models that learn directly from molecular representations, with a special focus on small molecules, peptides and proteins. Large language model-based chemistry assistants and agentic systems are excluded,[19,20] as the focus is on models that learn molecular representations directly rather than on systems that primarily leverage general-purpose language models and external tools. Nonetheless, the delineated criteria are broadly applicable.

The report that introduced the term 'foundational'[2] sets out five properties relevant to their development: (i) *expressivity*, the ability to represent real-world information; (ii) *scalability*, the capacity to accommodate increasingly high volumes of high-dimensional data; (iii) *multimodality*, the ability to process and integrate data from different sources and domains; (iv) *memory capacity*, the ability to retain and retrieve acquired information; and (v) *compositionality*, the ability to combined learned elements to address novel tasks. While these properties provide a conceptual basis for model design, they do not directly translate into operational criteria for assessing the foundational capabilities of a model. They are therefore used here as a conceptual vocabulary rather than as an evaluation checklist, with individual properties invoked whenever necessary.

**Figure 1. Proposed testable criteria for the 'foundational' nature of molecular models.** *Generality* asks whether a single model spans entity classes, property types and tasks; *transferability* asks whether it outperforms specialized or non-deep-learned models with minimal or no retraining; *out-of-distribution generalization* asks whether accuracy is stable under distribution shifts. They relate to desirable properties of a foundation model: multimodality (capacity to process content from different sources and domains), expressivity (capacity to capture and represent the training distribution flexibly), memory capacity (ability to store acquired knowledge and retrieve it for a new task), and compositionality (capacity to recombine what has been learned to handle situations absent from training). All three criteria rest on the same substrate: molecular representations, deep learning architectures, and the data and tasks used for training.

To assess foundational capabilities in the molecular sciences, three complementary criteria are proposed:

1. *Generality*: does a model work across different entities, properties and tasks? This relates to *multimodality*[2] and depends on molecular representations and model architecture.[21]
2. *Transferability*: can a model be reused on new tasks with minimal or no task-specific retraining, while outperforming simpler baselines? This relates to *expressivity* and *memory capacity*,[2] which enable the reuse of learned knowledge.
3. *Out-of-distribution generalization*: can a model extrapolate beyond its training distribution to novel chemical space? This relates to *compositionality*[2] and the ability to combine learned knowledge in new ways.

The three criteria are complementary in how they assess a model's capabilities: *generality* tests their breadth, *transferability* their reusability across tasks, and *generalization* their robustness beyond the training set distribution. Together, these criteria provide a conceptual framework for assessing the foundational nature of foundation models. They were applied to recent literature (Tables 1 and 2), with four of the five architectural properties mapped onto them (Fig. 1). Scalability is the exception, as the evidence assembled here suggests that in the molecular sciences model scale alone might not be the primary determinant of foundational capabilities.

The ultimate question is simple: judged by these three criteria, *are we there yet with foundation models in the molecular sciences?*

**Table 1. Selected Large-Scale Models in Chemistry and Structural Biology, provided as a context for the reader.** Selected models were grouped by application domain and reported for the type of molecules they can deal with and the properties they were applied to. An extensive survey of the state-of-the-art can be found elsewhere.[6] (Abbreviations: ADMET = Absorption, Distribution, Metabolism, Excretion, and Toxicity; LLM = Large Language Model; MD = Molecular Dynamics; SMILES = Simplified Molecular Input Line Entry System; 2D = Two-dimensional; 3D = Three-dimensional.)

| Application Domain | Selected Models | Molecule Type(s) | Property Type(s) |
|---|---|---|---|
| Protein structure and complex prediction. | AlphaFold2[9]; AlphaFold3[11]; Boltz-1[13]; Boltz-2[12]; RoseTTAFold All-Atom[22]; ESMFold.[23] | Proteins; protein complexes; nucleic acids; small-molecule ligands (AlphaFold3, Boltz); covalent modifications and metal ions (RoseTTAFold All-Atom). | 3D structure prediction; binding-affinity prediction (Boltz-2). |
| Protein conformational dynamics / ensemble generation. | BioEmu.[24] | Proteins (single-chain monomers). | Conformational ensemble sampling; free energy / thermodynamic stability estimation. |
| Protein sequence representation and design. | ESM-2[23]; ESM-3;[14] ESMC.[25] | Proteins. | Sequence embeddings; contact/structure prediction; de novo generative protein design; large-scale sequence annotation and similarity search (ESMC). |
| Peptide representation. | PepBERT[26]; PepLand[27]; HELM-BERT.[28] | Peptides: canonical (PepBERT); non-canonical amino acids (PepLand); cyclic and chemically modified peptides (HELM-BERT). | Bioactivity classification; membrane permeability; protein-peptide binding. |
| Small-molecule property prediction. | ChemBERTa-2[29]; MolFormer-XL[30]; Uni-Mol[31]; CLAMP[32]; CheMeleon[33]; Mol-JEPA;[34] MolGPS;[35] MiniMol.[36] | Small molecules (SMILES strings, 2D/3D molecular graphs; descriptors). Additional inputs: natural-language assay descriptions (CLAMP), and assay measurements, cellular phenotypes and pretrained embeddings (Mol-JEPA). | Bioactivity / ADMET property prediction; virtual screening; Supervised multitask learning on ~3,300 labelled tasks (MolGPS and MiniMol). |
| Materials and interatomic potentials. | M3Gnet;[37] CHGNet;[38] MACE-MP-0[39]; MatterSim;[40] GNoME;[15] eSEN-omol;[41] ANI-2x;[42] AIMNet2.[43]9/23/26 7:11:00 PM | Inorganic crystals; organic molecules and small proteins (MACE-MP-0); enzymes with bound substrates in solvent (eSEN-omol); drug-like organic molecules, neutral and charged molecules (ANI-2x; AIMNet2). | Energies and forces; structural relaxation; MD stability; discovery of stable materials; enzymatic reaction barriers (eSEN-omol). |
| Cross-domain / multi-entity pretraining. | JMP-1[44]; ESM-AA[45]; UMA.[46] | Small organic molecules + catalytic surfaces (JMP-1); proteins + atomic-resolution small molecules (ESM-AA); small molecules, inorganic materials and catalysts (UMA). | Atomic property prediction across chemical domains; joint protein-ligand tasks; transfer without fine-tuning across molecules, materials, and catalysts (UMA). |
| Cross-domain generative models. | NatureLM;[47] UniGenX.[48] | Small molecules, proteins, DNA, RNA, materials (NatureLM); small molecules, proteins, inorganic crystals (UniGenX). | Text-instructed generation and optimization; cross-domain generation; structure prediction; property prediction. |
| Multimodal molecule-text. | MoMu[49]; MoleculeSTM.[50] | Small molecules + natural-language text. | Molecule captioning; cross-modal retrieval; text-guided molecule editing. |

## Molecular Foundation Models through the Lenses of Generality, Transferability and Generalization

### 1. Generality across properties and domains

Logic. Generality can be understood as the ability of a single model to operate across different classes of molecular entities (e.g., small molecules, peptides, proteins, materials), rather than being restricted to a single domain. It also extends to the range of properties a model can capture (e.g., physico-chemical, structural and biological) and tasks it can perform (e.g., de novo design and property prediction). These dimensions are interdependent and shaped by the choice of data, molecular representation and model architecture, which together determine what problems a model can address.[21]

State of the art. Generality is tightly connected to the chosen molecular representation. Small-molecule models (e.g., ChemBERTa-2[29] and MolFormer-XL[30], Table 1) are usually pretrained on molecular strings[51] or molecular graphs, whose atom-level representations can become computationally inefficient for proteins containing thousands of atoms. Conversely, protein language models (e.g., ESM-2[23]; ESM-3[14]) operate on amino-acid sequences and lack native representations for small molecules. More general representations can address this divide. For instance, ESM-AA[45] unifies residue- and atom-level tokens into a single vocabulary, enabling a protein language model to process small molecules. Peptides occupy a middle ground (often overlooked) between small molecules and proteins, often falling outside the scope of protein models restricted to natural amino acids, and challenging small-molecule models with their long, polymer-like sequences.[52] Peptide-specific models (PepBERT,[26] PepLand,[27] HELM-BERT[28]) address this gap but remain separate from large-scale protein or small-molecule ecosystems.

Representation choice also constrains generality across properties. While molecular strings and two-dimensional graphs are in principle applicable to any endpoint, properties governed by three-dimensional and electronic structure (e.g., conformational energies, reactivity, and spectra) often require additional information or specialized model families,[21] such as interatomic potentials.[53] Representation-dependent limitations can arise even within a single entity class: structural information learned from SMILES[51] strings does not necessarily transfer to graph neural networks. Such findings suggest that pretraining on one representation might itself not be general enough to transfer to a different representation of the same molecule.[49]

Generality across different molecular entities has been shown, albeit with restricted task domains. JMP-1[44] spans small organic molecules and catalytic surfaces, yielding a 59% average improvement over training from scratch, although its scope remains atomic property prediction. Materials-focused interatomic potentials such as MACE-MP-0[39] cover solids, liquids, gases, and even the dynamics of a small protein, but remain confined to a shared atoms-and-forces formalism rather than spanning distinct task types.

Biomolecular structure prediction provides one of the clearest examples of generality across molecular entities (Table 1). AlphaFold3[11] predicts the structures of complexes containing proteins, nucleic acids, small-molecule ligands, ions, and modified residues, outperforming specialized tools according to the authors' benchmarks. Its mixed molecular representation – describing proteins and nucleic acids at the residue level and ligands at the atom level – avoids the size problem discussed above. Boltz-2[12] extends the scope by predicting protein-ligand binding affinity alongside structure, thereby covering more than one task type. Nonetheless, both models remain centered on the structure of molecular complexes (and the corresponding affinity), rather than a broader range of tasks. BioEmu[24] illustrates a different trade-off, by predicting protein conformational ensembles and folding free energies, with mean absolute errors below 1 kcal $mol^{-1}$ against both molecular dynamics and experiments. This thermodynamic task is, however, gained by restricting its scope to single protein chains at one temperature.

Recent models aim to extend generality across both molecular entities and tasks. NatureLM[47] represents small molecules, proteins, DNA, RNA, and materials via sequences of a common language, steered by text instructions, whereas UniGenX[48] combines an autoregressive transformer with a diffusion head to jointly generate molecular sequences and coordinates. Both span multiple entity classes and integrate generation and prediction within a single model. However, to date, their performance relative to specialized baselines has yet to be extensively and independently validated.

Verdict. Generality remains partial in the molecular domain, with no model yet demonstrating generality across molecular entities, properties, and tasks. Models spanning multiple molecular entities are typically restricted to a narrow range of properties and tasks, and vice versa. Recent unified generative models aim to overcome this limitation, but their broader capabilities await independent validation against specialized baselines. Molecular representation thus emerges as a key bottleneck, with approaches combining different levels of resolution offering a promising direction for future development.

**2. Transferability to downstream tasks**

Logic. Transferability asks whether a model, used off the shelf or lightly fine-tuned, outperforms simpler, task-specific baselines. This criterion tests whether pretraining captures knowledge that is reusable beyond the original training objective. This is particularly relevant in chemistry, where labeled data are often scarce and expensive to obtain, making the ability to outperform specialized models particularly valuable in low-data regimes. Transferability also offers an indirect test of scalability, revealing if and how increasing the pretraining data or model size improves the downstream performance.

State of the art. Assessing transferability requires specifying the baseline, the extent of task-specific fine-tuning, and the evaluation benchmark. In small-molecule property prediction, a common baseline[54–56] consists of molecular fingerprints (e.g., Extended Connectivity Fingerprints [ECFPs][57]) which encode local molecular substructures. An analysis spanning 25 pretrained embedding models across 25 benchmark datasets found negligible or no improvement over a simpler fingerprint-based model for nearly all neural models.[56] Only CLAMP (Contrastive Language-Assay Molecule Pre-Training)[32], itself built using fingerprints, performed significantly better. Similarly, when used as frozen embeddings with an otherwise identical classifier, MolFormer-XL[30] and ChemBERTa-2[29] underperformed ECFP baselines on a single screening task,[58] consistently with broader evaluations of large pretrained models.[59]

CheMeleon,[33] pretrained to predict classical descriptors, outperforms random forest and graph-based models, although, like most models tested, it struggles with activity cliffs.[33,54] Mol-JEPA,[34] learns from molecular descriptors, assay measurements and embeddings from other pretrained models, reports lower average errors than an ECFP random-forest baseline across ADMET benchmarks, with the largest improvements on smaller datasets.[34] The evidence of CLAMP, CheMeleon and Mol-JEPA suggest that incorporating established cheminformatics knowledge may be more effective than relying exclusively on 'raw' molecular structures. A comparison of five molecular foundation models with classical baselines for ADMET prediction supports this reading:[60] no model performed best consistently, and in the low-data scenarios the strongest results came from a pretrained tabular model (TabPFNv2[61]) operating on precomputed descriptors. Similarly, fingerprint-based descriptors have been shown to match or exceed graph neural networks on peptides function prediction, suggesting the problem is not specific to small molecules.[62]

Other approaches have focused on expanding the pretraining supervision: MolGPS[35] and MiniMol[36] are pretrained on large multi-task label mixtures. While MolGPS improves with the model size and the number of pretraining labels,[35] MiniMol achieves comparable downstream performance with two orders of magnitude fewer parameters.[36] This suggests that scale might improve transferability within a model family or architecture, without necessarily explaining performance differences across distinct approaches.

Transferability has been evaluated less systematically for materials-focused potentials. These models approximate expensive quantum-mechanical energies and forces and are typically pretrained on large datasets generated via standard density functional theory (DFT) approximations. Assessing transferability in interatomic potentials requires distinguishing between their application to previously unseen atomic systems, their adaptation to different levels of theory, and their use in downstream tasks beyond energy and force prediction. Matbench,[63] which evaluates thirteen models on crystal stability prediction, finds that universal potentials outperform fingerprint baselines. A different test is adaptation to a more accurate level of theory with limited additional data: CHGNet,[38] fine-tuned on 1,000 structures, outperformed a model trained from scratch on more than 10,000 structures,[64] but only after refitting the elemental reference, indicating that the benefits might depend on choices that lie outside the model pretraining. UMA[46] and eSEN-omol,[41] both drawing on OMol25[65] provide examples of transfer without task-specific fine-tuning: UMA matched or exceeded domain-specialized models on quantum-chemical tasks relevant to drug discovery, while eSEN-omol reproduced experimental barriers and mechanistic features in three unrelated enzymes in explicit solvent, without system-specific tuning. A complementary approach narrows the chemical domain rather than expanding it: potentials designed for the elements present in drug-like organic matter report accuracy comparable to that of the DFT methods they replace and higher than that of semi-empirical methods (their natural non-deep-learned baselines).[42,43]

Structure prediction models such as AlphaFold3[11] and Boltz-2[12] require a slightly different interpretation of transferability. Rather than being pretrained and then adapted, they are mostly used as they are, so their transferability lies in whether their outputs remain useful for tasks beyond structure prediction. Without any fine-tuning, Boltz-2 affinity predictions were reported to approach the accuracy of physics-based free energy calculations with 1000 times lower computational cost, and to outperform the top-ranking participants in the CASP16 affinity challenge in a retrospective evaluation.[12] An independent evaluation on two drug targets, however, found only weak to moderate correlations between its predictions and physics-based binding free energy estimates.[66] The two models with the widest entity coverage (NatureLM[47] and UniGenX[48]) do not resolve the question: their evaluations involve extensive task-specific supervised adaptation, making it difficult to assess how much of their downstream performance is attributable to transferable pretrained representations rather than task-specific adaptation.

Verdict. Transferability with little or no task-specific adaptation remains difficult to establish consistently across molecular domains. In small-molecule property prediction, pretrained models often fail to outperform simple fingerprint-based baselines, with competitive examples such as CLAMP,[32] CheMeleon[33] and Mol-JEPA[34] incorporating established chemical knowledge into their representations or pretraining objectives. In atomistic modeling, UMA[46] and eSEN-omol[41] provide examples of downstream applications without task- or system-specific fine-tuning. By contrast, evaluations of models spanning multiple molecular entity classes[47,48] rely on separate task-specific adaptation, making the contribution of transferable pretrained representations difficult to isolate. Overall, claims of transferability are difficult to assess without comparisons against appropriate simple baselines and an explicitly stated adaptation budget.

### 3. Out-of-distribution generalization

The logic. Out-of-distribution (OOD) generalization asks whether a model remains accurate on chemical systems that differ systematically from those present in the training data. It is closely related to *compositionality*,[2,67] the ability to recombine learned components to address novel situations. Failure may arise because the relevant components were not learned, because the model cannot recombine them effectively, or because its inference procedure fails to recover an otherwise accessible prediction. Distinguishing these possibilities is important when interpreting model performance. Unlike the other two criteria, OOD generalization also depends critically on how novelty is defined: molecular similarity is, to some extent, "in the eye of the beholder",[68] and data splits based on molecular scaffold, property range, protein family, or sequence similarity probe different forms of distribution shift.[69–71]

State of the art. OOD generalization has received considerable attention in small-molecule modeling, especially in terms of how it is evaluated. Scaffold splits have been shown to overestimate virtual screening performance,[72] while standard splits might overestimate the generalization of protein-ligand affinity predictors to novel protein families.[73–75] More stringent evaluations reveal further limitations: under perimeter splits – which assign molecules at the outskirts of the distribution to the test set – all models examined lost predictive accuracy relative to random splits, with standard graph neural networks degrading more than pretrained models.[60] The BOOM benchmark,[70] which evaluated more than 140 model-task combinations under distribution shift, found that no model generalized well across tasks. Even the best-performing model had an average OOD error approximately three times higher than its in-distribution error. The authors concluded that pretraining primarily improves interpolation rather than extrapolation.[70] This echoes the broader machine-learning literature on distribution shift, which identifies molecular property prediction as a recurring challenge.[76] Related limitations extend to multimodal language models applied to chemistry and materials, which, despite strong pattern-recognition and information-retrieval capabilities, struggle with applications to novel situations.[77]

Improvements have been reported from reformulating the prediction task, rather than increasing the scale. Predicting a candidate's property relative to a known reference rather than its absolute value can improve accuracy and data efficiency,[78,79] and has improved extrapolation in the settings examined.[71]

Machine-learned interatomic potentials also exhibit limitations under distribution shift. In finite-temperature molecular-dynamics simulations of minerals, two widely used potentials showed mean density deviations of approximately 40% and 80%[80] – error ranges that may compromise the resulting molecular-dynamics trajectories. In a separate study, a potential trained on graphene oxide was 'zero-shot' transferred to isolated small molecules and reactions, revealing measurable performance gaps.[81] The authors concluded that transfer beyond the training data domain must be measured case by case.[81,82]

Protein-ligand cofolding models exhibit a similar pattern. AI-based docking methods frequently generate structures that fail basic physical checks,[83–85] with failure rates increasing with protein sequence dissimilarity.[83,84] In a benchmark of 2,600 protein-ligand complexes,[84] accuracy declined markedly for complexes dissimilar to training data. The binding pocket was often modelled correctly even when the ligand pose was not, suggesting that the difficulty lies in predicting protein-ligand interactions rather than protein structure alone; this has been interpreted as evidence of memorization of ligand poses.[84] Cofolding methods still outperform conventional docking on average, but remain challenged by novel binding poses and by multi-ligand targets.[86] Frequent violations of basic physical and chemical principles point in the same direction, namely overfitting to features of the training corpus rather than learning transferable interaction physics.[87] These observations echo a well-documented behavior of deep networks, which can memorize individual training instances instead of learning the abstract, reusable regularities that support generalization[88,89] and compositional recombination.[67] Some evidence shows that part of this gap is addressable without a more general model: training-free strategies that leave the weights untouched[90,91] – perturbing the conditioning, steering generation with physical restraints, or sampling more broadly and re-ranking – partially recover accuracy on structurally less similar targets, suggesting the failure is partly one of sampling rather than of representation only.

Evidence of OOD generalization also exists in protein conformational modeling. A machine-learned coarse-grained force field[92] trained on a limited set of protein domains recovered the folded states of proteins with low sequence similarity to its training data. It also captured folding upon binding of a disordered peptide despite the lack of protein complexes in its training data.[92] BioEmu[24] retained its accuracy on proteins sharing no more than 40% sequence identity with its training set. However, its performance degraded where the training data was sparse, recovering cryptic pockets more reliably in their ligand-bound structures rather than in their unbound counterparts.[24]

Verdict. OOD generalization remains a major challenge across molecular domains. Performance on conventional benchmarks does not necessarily predict accuracy under shifts in molecular structure, protein sequence, or conformational space, making the definition of novelty and the design of evaluation splits central to any claim of generalization. The evidence also shows that poor OOD performance may not arise solely from limitations of the learned representation, since task formulation and inference-time strategies can improve predictions without increasing model scale or retraining. At the same time, the examples of successful generalization in protein conformational modeling show that extrapolation is possible, although its limits remain closely tied to the coverage of the training data.

**Table 2. Representative model families analyzed via the three criteria.** Each family is evaluated on generality across chemical entities and across task types, on transferability (with the adaptation budget under which it was established and the baseline it was compared against), and on out-of-distribution generalization (with the type of split used). For each criterion we define the following labels: "●" = criterion met, "◐" = criterion partially met; "○" = criterion not met; "–" = criterion not independently tested against a suitable baseline. Letters in square brackets indicate additional notes by the author.

| Model family | Generality | | Transferability | OOD generalization |
|---|---|---|---|---|
| | Across entities | Across tasks | Adaptation budget · baseline | Split type |
| **Biomolecular structure & complex prediction**<br>AlphaFold3, Boltz-2 | ● proteins, nucleic acids, ligands, ions[11] | ◐ structure; affinity (Boltz-2 only)[12] | ◐ zero-shot · vs physics-based FEP / docking[12,66] | ○ sequence / pocket similarity split[83,84,86,87] |
| **Protein language models**<br>ESM-2, ESM-3, ESMC | ○ proteins only[14,23,25] | ◐ embeddings, structure, design[14,23] | – no non-deep-learned baseline [a] | – [a] |
| **Peptide representation**<br>PepBERT, PepLand, HELM-BERT | ○ peptides only[26–28] | ◐ bioactivity, permeability, binding[26–28] | – not independently tested against a non-deep-learned baseline | – |
| **Small-molecule embedding models**<br>ChemBERTa-2, MolFormer-XL, Uni-Mol | ○ small molecules only[29–31] | ◐ any 2D endpoint; 3D / electronic left to other families[21] | ○ frozen · vs ECFP + random forest[56,58] | ○ scaffold / property-range splits; activity cliffs[54,69,70] |
| **Knowledge-injected small-molecule models**<br>CLAMP, CheMeleon, Mol-JEPA | ○ small molecules only[32,33] | ◐ bioactivity / ADMET[32,33] | ● frozen or light fine-tuning · vs ECFP + random forest[33,56] | ○ activity cliffs[33,54] |
| **Machine-learned interatomic potentials**<br>MACE-MP-0, CHGNet, UMA, eSEN-omol | ● solids, liquids, molecules, proteins, enzymes[39,41,46] [b] | ○ energies and forces only[53] | ● zero-shot · vs Voronoi-fingerprint random forest, [63] classical force fields[41,46,64] | ○ perturbed structures; zero-shot to new systems[80–82] |
| **Cross-domain pretraining**<br>JMP-1, ESM-AA | ◐ molecules + catalytic surfaces; proteins + ligands[44,45] | ○ atomic property prediction (JMP1), protein-ligand tasks (ESM-AA)[44,45] | ◐ fine-tuned · vs training from scratch[44][c] | – |
| **Cross-domain generative models**<br>NatureLM, UniGenX | ● molecules, proteins, nucleic acids, materials[47,48] [d] | ● generation + prediction[47,48][d] | – fine-tuned per task; a single set of weights never evaluated[47,48] | – |

[a]Non-deep-learned baselines for protein fitness prediction exist but are not discussed here. [b]Broad coverage within a single atoms-and-forces formalism, not across task types. [c]Compared with training from scratch rather than with a non-deep-learned baseline. [d]Developers' claims, not yet independently validated.

## Conclusions and outlook

The analysis presented here is not intended as a scorecard, but as a map of demonstrated capabilities and untapped opportunities (Table 2). Several of the models surveyed represent important scientific achievements. Biomolecular structure prediction[11] and machine-learned interatomic potentials[53] illustrate the value of established data infrastructures, as well as relatively well-defined target properties and tasks. Structure predictors benefit from decades of community curation in public structural archives and from evolutionary information encoded in sequence databases; interatomic potentials benefit from quantum-mechanically defined reference energies and forces. These advantages are less readily available for properties such as bioactivity and metabolism, where experimental measurements may vary across assays and conditions.[93–97] Architectural advances alone cannot fully compensate for these differences. A central lesson is therefore that progress toward molecular foundation models depends not only on model development, but also on the community effort required to build coherent, well-characterized datasets.

This analysis also identifies molecular representation as a major constraint on generality across entity classes. Two directions are being explored: mixed-resolution representations[11,22,45] that describe different entities at appropriate levels of detail (e.g., proteins by residue and ligands by atom), and unified representations[47,48] that encode diverse molecular entities within a common sequence language. Whether either approach supports broader generalization remains an open, testable question. Peptides offer a useful test case because they sit at the interface between small molecules and proteins and challenge models to connect atomistic and sequence-based descriptions.[52]

A further question concerns the definition of 'novelty'. OOD evaluations typically define novelty externally, using criteria such as molecular scaffold, property range, or protein family. Yet a model's learned representation may organize chemical space differently.[98] Disagreement between these perspectives could be informative: some compounds considered novel by a chemist may lie in regions familiar to a model, while the reverse may also occur.[98] Identifying such discrepancies could reveal where chemical novelty and predictive accuracy can coexist.

The three criteria are only as informative as the evaluations used to test them. Priorities for future benchmarking include evaluations spanning molecular entity classes, property types, and task types; distribution-shift tests reported alongside conventional splits; explicit adaptation budgets that distinguish off-the-shelf use from extensive fine-tuning; and comparisons with appropriate simple, task-specific baselines. Structured evaluation frameworks have already been advocated for foundation models more broadly,[99] and related diagnostic efforts are emerging for interatomic potentials.[100] A shared evaluation vocabulary across the molecular sciences would make results accumulate rather than be re-discussed domain by domain, as well as promote the cross-contamination between 'siloed' communities that currently use different benchmarks and conventions.

The evidence surveyed also identifies opportunities beyond increasing model or dataset size. Examples include incorporating established chemical knowledge through fingerprints, molecular descriptors, and assay information;[32,33] designing physically grounded coarse-grained models that generalize to proteins with low sequence similarity to their training data;[92] reformulating property prediction in terms of differences from rather than absolute values;[71] and improving inference through sampling, physical restraints, or re-ranking without retraining the underlying model.[90,91] These approaches do not establish that prior knowledge is always preferable to scale, but they show that the choice of representation, objective, and inference procedure can materially affect downstream performance. In machine learning at large, the expectation has been that general methods riding on data computation eventually overtake hand-crafted knowledge.[101,102] Chemistry offers a particularly useful setting in which to examine where that crossover lies: its prior knowledge is well codified, with molecular representations themselves being a model of a far more complex physical reality.[103] This knowledge has been accumulated over centuries and can be built into models and tasks rather than being re-learned from data.

Finally, the three criteria examined here are *necessary but not sufficient*. Chemistry is also a design discipline: its questions concern not only the properties of existing systems, but also the consequences of changing them and deciding what to make next. Accurate prediction from observational data does not, by itself, establish reliable reasoning about the effects of an intervention.[104] Chemical common sense, causal reasoning,[104] and calibrated uncertainty[105,106] are therefore important directions for future work, provided that each can be translated into explicit, testable capabilities. These questions may also provide common ground between molecular foundation models and chemistry assistants built on general-purpose language models, which are increasingly evaluated on related capabilities.[107,108] The growing reliance on a small number of widely reused models could make their limitations more widely shared,[2] reinforcing the importance of maintaining a diversity in models, representations, and research directions.

Ultimately, *'foundational'* should not be treated as a label assigned in advance, but as a hypothesis about a model's capabilities. Testing that hypothesis against explicit criteria can clarify which molecular problems already admit broadly reusable solutions, which capabilities remain limited, and where further progress is needed.

**Box 1 | Glossary of selected terms.**

**ADMET** (Absorption, Distribution, Metabolism, Excretion, and Toxicity). The pharmacokinetic and safety properties that determine whether a bioactive molecule is also a viable drug candidate, as distinct from its activity against the target.

**Activity cliff.** A pair of structurally similar molecules whose biological activities differ markedly.[109]

**CASP** (Critical Assessment of Structure Prediction). A community effort, in which structure prediction methods are assessed blind on targets whose experimental structures are not yet public.[110]

**Co-folding.** The joint prediction of the three-dimensional structure of a protein and of its bound ligand within a single model, as opposed to placing a ligand into a fixed protein structure, as in conventional docking.

**Distribution shift.** A difference between the data a model was trained on and the data to which it is applied; a general concern in machine learning, not specific to chemistry.[76]

**ECFP** (extended-connectivity fingerprint). The most widely used family of *molecular fingerprints*, in which circular atom environments of increasing radius are enumerated and hashed into a fixed-length vector.[57]

**Fine-tuning.** The continued training of a pretrained model on data from a specific task, typically with far fewer examples than used during pretraining (*see* Transfer learning).

**Interatomic potential.** A model that predicts the energy of a system of atoms, and the forces acting on them, as a function of atomic positions, used as a fast surrogate for quantum-mechanical calculations in molecular simulations.

**Molecular descriptor.** Numbers or vectors capturing pre-defined chemical information.[111]

**Molecular fingerprint.** A vector encoding the presence of predefined substructures in a molecule.

**Molecular graph.** A representation of a molecule in which atoms are nodes and bonds are edges, each annotated with chemical features; it is the natural input of graph neural networks.[21]

**Molecular representation.** The encoding through which a molecule is presented to a model, for example a character string (such as SMILES[51]), a two-dimensional graph, or a set of three-dimensional coordinates.

**Molecular scaffold.** The core structural frameworks of molecules, excluding their peripheral substituents.

**Pretraining.** The initial, largely task-agnostic training of a model on large and typically unlabeled data, preceding any adaptation to a specific task (*see* Transfer learning).

**SMILES** (Simplified Molecular Input Line Entry System). A string notation that encodes the 2D structure of a molecule.[51]

**Transfer learning.** The reuse of knowledge acquired on one task to improve performance on another. In its most common form, pretraining and fine-tuning are its two stages.

**Zero-shot.** The application of a model to a task without any task-specific training example.

## Acknowledgements

FG acknowledges inspiration and support from the European Research Council (ERC) under the European Union's Horizon Europe research and innovation programme (ReMINDER, grant agreement no. 101077879), and from the Dutch Research Council (NWO; grant VI.Vidi.233.164 of the Vidi ENW research programme, https://doi.org/10.61686/IVDFS18985). Views and opinions expressed are those of the author only and do not necessarily reflect those of the European Union or the European Research Council Executive Agency. Neither the European Union nor the granting authority can be held responsible for them. FG also thanks the Molecular Machine Learning team at TU/e, in particular Nopsinth (Will) Vithayapalert and Katarina Elez, for fruitful scientific discussions on structure prediction, foundation models, and their limitations.